\documentclass[aps,superscriptaddress,amsart]{revtex4}
\usepackage{graphics}
\usepackage{amssymb}
\usepackage{amsmath}
\usepackage{epsfig}
\usepackage{color}
\usepackage{verbatim}
\usepackage{float}
\usepackage{mathrsfs}
\usepackage{esint}
\usepackage{tabularx}
\usepackage{multirow}
\begin{document}



\title{Larkin-Ovchinnikov phases in two-dimensional square lattices.}

\author{J.E. Baarsma}
\author{P. T\"orm\"a}\thanks{paivi.torma@aalto.fi}
\address{COMP Centre of Excellence, Department of Applied Physics, Aalto University, FI-00076 Aalto, Finland}

\begin{abstract}
We consider a two-component gas of fermions in optical lattices in the presence of a population imbalance within a mean-field theory. We study phase transitions from a normal gas of unpaired fermions to a superfluid phase of Bose-condensed Cooper pairs. The possibility of Cooper pairs with a nonzero center-of-mass momentum is included, which corresponds to a so-called Fulde-Ferrel-Larkin-Ovchinnikov (FFLO) state. We find that for population imbalanced systems such states can form the ground state. The FF and LO state are compared and it is shown that actually the LO state is energetically more favorable. We complete the mean-field phase diagram for the LO phase and show that it is qualitatively in excellent agreement with recent diagrammatic Monte Carlo calculations. Subsequently, we calculate the atomic density modulations in the LO phase. 
\end{abstract}

\maketitle


\section{Introduction}
The field of ultracold atoms grew rapidly after the major breakthrough in 1995, the first realization of a Bose-Einstein condensate in a weakly interacting gas of alkali atoms \cite{Anderson14071995,PhysRevLett.75.1687,PhysRevLett.75.3969}.
Research on ultracold fermionic gases naturally followed and quantum degeneracy was achieved for the first time in 1999 in ultracold potassium \cite{DeMarco10091999}, after which a large number of exciting experiments have been performed. For instance, with the use of Feshbach resonances the BEC-BCS crossover was studied \cite{2008NCimR..31..247K,RevModPhys.80.885,RevModPhys.80.1215,zwerger2011bcs}. Furthermore, by inducing spin flips in a Fermi gas the influence of a population imbalance on the phase transition to the superfluid state was studied and it was found that the phase diagram is governed by a tricritical point \cite{Zwierlein27012006,Partridge27012006,2008Natur.451..689S}. Namely, for small imbalances a continuous transition was found from the normal to the BCS superfluid state, while for larger imbalances the transition is discontinuous, which shows up as a region of phase separation in the phase diagram.

The high tunability of ultracold Fermi gases make these systems a very promising platform for observing exotic states of matter, such as Fulde-Ferrel-Larkin-Ovchinnikov (FFLO) superfluid states, where the Cooper pairs have a nonzero center-of-mass momentum \cite{Fulde1964a,Larkin1964a,Larkin1965b}. Studies of an atomic gas in one spatial dimension have already been reported that are consistent with the existence of the FFLO state \cite{huletFFLO2010}, but a decisive experiment that observes the FFLO correlations has not been carried out yet. The difficulty of an experimental observation in homogeneous systems in three spatial dimensions are the extremely low transition temperatures in the weakly-interacting regime where FFLO phases are predicted to be present  and the smallness of the parameter regime for its existence \cite{PhysRevB.71.214504,PhysRevLett.95.117003,PhysRevLett.96.060401,PhysRevLett.96.110403,PhysRevA.75.063601,PhysRevB.76.184104,PhysRevA.82.013624}. For reviews on the search for the FFLO state in ultracold gases see \cite{2007AnPhy.322.1790S,2010RPPh...73g6501R,Gubbels2013255}.

The realization of ultracold atoms loaded into optical lattice potentials has opened up the possibility for studying a whole new class of many-body systems \cite{1998PhRvL..81.3108J,2002Natur.415...39G}. In recent experiments using fermionic atoms, low enough temperatures were achieved in the 3D Hubbard model to observe short-range anti-ferromagnetic correlations \cite{2013Sci...340.1307G,2015Natur.519..211H}. Another very promising development is the realization of the quantum gas microscope, initially for bosonic atoms \cite{2009Natur.462...74B,2010Natur.467...68S}, which allows the probing of strongly-interacting fermions at the single atom level \cite{PhysRevLett.114.193001,PhysRevLett.114.213002,2015NatPh..11..738H,2015arXiv151004744E}.

Theoretically, it has been predicted that optical lattices enhance the FFLO superfluid pairing, due to nesting of the Fermi surfaces \cite{PhysRevLett.99.120403,1367-2630-10-4-045014}. The parameter regime where FFLO states are stable is therefore larger in optical lattices \cite{PhysRevLett.99.120403,1367-2630-10-4-045014,PhysRevLett.100.116405,PhysRevLett.104.165302,
TempereFF3D1D_2011,PhysRevA.86.023630,2015JPCM...27v5601R}, making this promising systems to observe these exotic quantum states of matter.
Most of the theoretical studies include the {\it plane-wave FF ansatz} for studying finite momentum pairing, because of its relative simplicity. However, it has been shown that for {\it homogeneous} Fermi gases actually the LO standing-wave ansatz is energetically more favorable \cite{PhysRevB.71.214504,PhysRevA.75.063601,JildouenHenk_PRA2013}. Here, the aim is to study whether also in {\it lattices in two dimensions} the LO state is more favorable, which has been shown for lattices in three dimensions \cite{PhysRevLett.104.165302}. Notably, the LO phase leads to an inhomogeneous density of the gas, which can be an important experimental signature of this exotic state.
In this paper, we first explain what FFLO states are in some detail. Subsequently, we study finite momentum pairing in population imbalanced fermions in optical lattices, by calculating the full mean-field thermodynamic potential for the LO ansatz and show that the LO superfluid is indeed more favorable than the FF state. We calculate phase diagrams, also at fixed filling fraction and compare our mean-field results with state of the art diagrammatic Monte Carlo (diagMC) results \cite{DiagMC_Troyer2015}. We calculate atomic densities that show the signature of LO pairing \cite{JildouenHenk_PRA2013}. 

\section{FFLO}
We consider two-dimensional optical lattices, by which we mean a lattice strongly confined in one dimension, populated by fermionic particles in two (pseudo)spin states, $\uparrow$ and $\downarrow$. The fermions can hop between nearest-neighboring sites and they interact when on the same site. The Hamiltonian describing these systems  reads
\begin{align}
\hat{H}=-t\sum_{<i,j>}\sum_{\sigma\in\{\uparrow,\downarrow\}}\hat{c}_{i\sigma}^\dagger\hat{c}_{j\sigma}-\sum_{i;\sigma}\mu_\sigma \hat{c}_{i\sigma}^\dagger\hat{c}_{i\sigma}+U\sum_i\hat{c}_{i\uparrow}^\dagger\hat{c}_{i\downarrow}^\dagger\hat{c}_{i\downarrow}\hat{c}_{i\uparrow},
\end{align}
where $\hat{c}_{i\sigma}^\dagger$ creates a fermionic particle at site $i$ in spin state $|\sigma\rangle$. The hopping between nearest-neighboring sites is characterized by the hopping parameter $t$, while different densities for the two spin components are introduced by having different chemical potentials $\mu_\sigma$. The on-site interaction $U$ is taken attractive in this paper.
In order to later on be able to calculate the thermodynamic potential, we perform a mean-field approximation
\begin{align}
U\sum_i\hat{c}_{i\uparrow}^\dagger\hat{c}_{i\downarrow}^\dagger
\hat{c}_{i\downarrow}\hat{c}_{i\uparrow}\simeq
U\sum_i\left\{\hat{c}_{i\uparrow}^\dagger\hat{c}_{i\downarrow}^\dagger
\langle\hat{c}_{i\downarrow}\hat{c}_{i\uparrow}\rangle
+\langle\hat{c}_{i\uparrow}^\dagger\hat{c}_{i\downarrow}^\dagger\rangle
\hat{c}_{i\downarrow}\hat{c}_{i\uparrow}
-\langle\hat{c}_{i\uparrow}^\dagger\hat{c}_{i\downarrow}^\dagger\rangle
\langle\hat{c}_{i\downarrow}\hat{c}_{i\uparrow}\rangle\right\},
\end{align}
and introduce the Cooper pairs $\Delta_i=U\langle\hat{c}_{i\downarrow}\hat{c}_{i\uparrow}\rangle$. This yields for the total Hamiltonian
\begin{align}
\hat{H}=-t\sum_{<i,j>,\sigma}\hat{c}_{i\sigma}^\dagger\hat{c}_{j\sigma}-\sum_{i,\sigma}\mu_\sigma \hat{c}_{i\sigma}^\dagger\hat{c}_{i\sigma}+\sum_{i}\left\{\hat{c}_{i\uparrow}^\dagger\hat{c}_{i\downarrow}^\dagger\Delta_i
+\Delta_i^\dagger\hat{c}_{i\downarrow}\hat{c}_{i\uparrow}
-\frac{\Delta_i^\dagger\Delta_i}{U}\right\}.
\label{Hgen}
\end{align}
Up to this point the discussion is very general. The above steps actually apply to different lattice dimensions and can be easily generalized to the homogeneous case or to different trapping geometries. And most importantly, no particular superfluid phase is assumed yet, which is what we elaborate on next.

Usually in BCS theory, when there is an equal number of $\uparrow$ and $\downarrow$ particles, Cooper pairs consist of two particles in different spin states and with opposite momentum, ${\bf k}$ and $-{\bf k}$, such that the pairs have no net momentum. The resulting phase of Bose-Einstein condensed pairs is a homogeneous superfluid phase, in the sense that the Cooper pair density is position independent. To describe phase transitions from a normal state to a BCS superfluid, it suffices to take into account position-independent Cooper pairs, i.e., pairs without net momentum
\begin{align}
\Delta_x\equiv\Delta_0\hspace{3mm}\text{(BCS)},
\end{align}
where here $x$ is used to denote position. In the balanced case this homogeneous phase is expected, because the Fermi surfaces of the two spin components match. The first pair to form is between two particles at the respective Fermi surfaces and they thus carry the same momentum, since $k_{F\uparrow}=k_{F\downarrow}$.

A question posed by Fulde and Ferrel and independently by Larkin and Ovchinnikov in the context of superfluid films in strong magnetic fields is, what happens if there is a mismatch between the Fermi surfaces of the two spin components? More specifically, what happens to the phase transition from the normal to the superfluid state in the presence of a population imbalance between the $\uparrow$ and $\downarrow$ particles? A Cooper pair formed between two particles at the two Fermi surfaces now carries a net momentum equal to $k_{F\uparrow}-k_{F\downarrow}$ and the resulting superfluid phase might be a more exotic superfluid phase than the homogeneous BCS state.
To study such superfluid phases, Cooper pairs carrying a net momentum are taken into account. In this paper, we focus on the Fulde-Ferrell (FF) and the Larkin-Ovchinnikov (LO) superfluid, that correspond to the following ansatzes for the Cooper pairs
\begin{align}
\Delta_x\equiv\Delta_0 e^{i{\bf q\cdot x}}\hspace{3mm}\text{(FF)}\label{FFCpair},\\
\Delta_x\equiv\Delta_0\cos({\bf q\cdot x})\hspace{3mm}\text{(LO)}\label{LOCpair}.
\end{align}
In both the FF and the LO phase the Cooper pairs have a net momentum ${\bf q}$. The difference is that in the LO phase translational symmetry is broken, while in the FF phase it is not \cite{2007AnPhy.322.1790S,2010RPPh...73g6501R,Gubbels2013255}, in addition to the $U(1)$ symmetry broken by $\Delta_0$. Next, we explain in some detail how to calculate the mean-field thermodynamic potential including the possibility of an instability towards the FF state. Implementing the LO state is in principle the same, but way more involved.

In order to calculate the thermodynamic potential $\Omega_\text{FF}$ describing phase transitions from the normal state to the FF superfluid state, we insert the FF Cooper pair ansatz Eq.(\ref{FFCpair}) into the Hamiltonian Eq.(\ref{Hgen}) and subsequently Fourier transform, which yields
\begin{align}
\hat{H}=\sum_{{\bf k},\sigma}(\varepsilon_{\bf k}-\mu_\sigma)\hat{c}_{{\bf k}\sigma}^\dagger\hat{c}_{{\bf k}\sigma}+\sum_{\bf k}\left\{\hat{c}_{{\bf k}\uparrow}^\dagger\hat{c}_{{\bf q-k}\downarrow}^\dagger\Delta_0+\Delta_0\hat{c}_{{\bf q-k}\downarrow}\hat{c}_{{\bf k}\uparrow}\right\}-\frac{\Delta_0^2}{U},
\label{hamnd}
\end{align}
where the dispersion $\varepsilon_{\bf k}$ depends on the lattice geometry, see Eq.(\ref{dispersion}) for the square lattice. The summation over ${\bf k}$ above is over the first Brillouin zone. We diagonalize the above Hamiltonian,
\begin{align}
\nonumber\hat{H}&=\sum_{\bf k}\left\{ \varepsilon_{\bf q-k}-\mu_\downarrow+(\hat{c}_{{\bf k}\uparrow}^\dagger,\hat{c}_{{\bf q-k}\downarrow})\left(\begin{array}{cc}
\varepsilon_{\bf k}-\mu_\uparrow&\Delta_0\\
\Delta_0&-(\varepsilon_{\bf q-k}-\mu_\downarrow)\end{array}\right)
\left(\begin{array}{c}
\hat{c}_{{\bf k}\uparrow}\\ 
\hat{c}_{{\bf q-k}\downarrow}^\dagger
\end{array}\right)
\right\}-\frac{\Delta_0^2}{U}\\
&=\sum_{\bf k}\left\{ \varepsilon_{\bf q-k}-\mu_\downarrow+(\hat{d}_{{\bf k,q}\uparrow}^\dagger,\hat{d}_{{\bf k,q}\downarrow})\left(\begin{array}{cc}
\hbar\omega_{{\bf k,q}\uparrow}&0\\
0&-\hbar\omega_{{\bf k,q}\downarrow}\end{array}\right)
\left(\begin{array}{c}
\hat{d}_{{\bf k,q}\uparrow}\\ 
\hat{d}_{{\bf k,q}\downarrow}^\dagger\end{array}\right)\right\}-\frac{\Delta_0^2}{U}\label{hambog},
\end{align}
where a unitary Bogoliubov transformation was made in the second line \cite{PhysRevA.82.013624}, introducing the quasi-particle operators $d_{{\bf k,q}\sigma}$ that are related to the particle operators via the coherence factors $u_{\bf k,q}$ and $v_{\bf k,q}$, similar to the usual BCS theory. The dispersions in the diagonalized Hamiltonian read
\begin{align}
\hbar\omega_{{\bf k,q}\sigma}=\sqrt{\left[\frac{\varepsilon_{\bf k}+\varepsilon_{\bf q-k}}{2}-(\mu_\uparrow+\mu_\downarrow)\right]^2+|\Delta_0|^2}\pm\left[\varepsilon_{\bf k}-\varepsilon_{\bf -k+q}-(\mu_\uparrow-\mu_\downarrow)\right],
\end{align}
with the $+$($-$) sign corresponding to the $\uparrow(\downarrow)$ quasi-particles. From this diagonalized Hamiltonian the partition sum $Z=\text{Tr}[\exp(-\beta \hat{H})]$ can now be calculated, where $\beta=1/k_\text{B}T$. Subsequently, the thermodynamic potential $\Omega_\text{FF}$ can be calculated via $\Omega=-\ln Z/\beta$, which results in
\begin{align}
\Omega_\text{FF}(\Delta_0,{\bf q})=\sum_{\bf k}\left\{\varepsilon_{\bf q-k}-\mu_\downarrow-\hbar\omega_{{\bf k,q}\downarrow}-\frac1{\beta}\sum_\sigma\ln\left(1+e^{-\beta\hbar\omega_{{\bf k,q}\sigma}}\right)\right\}-\frac{\Delta_0^2}{U}.
\end{align}
The FF thermodynamic potential depends on both the amplitude $\Delta_0$ and the Cooper pair momentum ${\bf q}$ and has to be minimized with respect to both parameters in order to find the ground state of the system for given parameters. The FF thermodynamic potential reduces to the BCS thermodynamic potential if ${\bf q}$ is taken to zero.

To caculate the LO thermodynamic potential one needs to insert the LO Cooper pair ansatz Eq.(\ref{LOCpair}) into the Hamiltonian Eq.(\ref{Hgen}). The part describing pair creation and annihilation reads after Fourier transforming
\begin{align}
\hat{H}_\text{pairs}=\sum_{\bf k}\left\{\hat{c}_{{\bf k}\uparrow}^\dagger\hat{c}_{{\bf q-k}\downarrow}^\dagger\Delta_0
+\hat{c}_{{\bf k}\uparrow}^\dagger\hat{c}_{{\bf -q-k}\downarrow}^\dagger\Delta_0
+\Delta_0\hat{c}_{{\bf q-k}\downarrow}\hat{c}_{{\bf k}\uparrow}
+\Delta_0\hat{c}_{{\bf -q-k}\downarrow}\hat{c}_{{\bf k}\uparrow}\right\},
\end{align}
from which it can be seen that the LO state is quite different from the BCS and FF states, since every fermion now couples to two other fermions, instead of to one. Namely, an $\uparrow$ fermion with momentum ${\bf k}$ now couples to both a $\downarrow$ fermion with momentum ${\bf -k+q}$ and to one with momentum ${\bf -k-q}$. This has huge consequences for calculating the thermodynamic potential. Namely, before we were able to rewrite the Hamiltonian in Eq.(\ref{hamnd}) using matrix multiplication with a $2\times2$ matrix, while currently a matrix with infinite dimensions would be needed. Instead of working with an infinite matrix, we start with a matrix of dimension $D=6$ to calculate the LO thermodynamic potential and increase its size until $\Omega_\text{LO}(\Delta_0,{\bf q})$ has converged. Close to a continuous phase transition, when $\Delta_0$ is small, the convergence is quite fast, whereas inside the LO superfluid phase larger matrices are needed.

In both the FF and LO state discussed above, the thermodynamic potential has to be minimized with respect to both the amplitude $\Delta_0$ and the pair momentum ${\bf q}$. The latter has both a magnitude and a direction. In the homogeneous case all directions of ${\bf q}$ are equivalent and only its magnitude matters for the thermodynamic potential. In the case of a lattice potential there is a difference and the energy for the different directions of ${\bf q}$ has to be compared to determine the ground state of the system, which we do below for the square lattice.

\section{Two-dimensional Square Lattice}
In this section we present results on the two-dimensional square lattice. We first compare for some fixed temperature and interaction strength the phase diagram for the FF and LO state and show that the LO state is actually more favorable. Consequently, we compare different directions of the broken translational symmetry for the LO state and show that the momentum along a lattice axis is most favorable. We then present some more phase diagrams for the LO state, both as a function of chemical potentials and as a function of temperature and polarization. We compare our phase diagram at fixed filling to recent diagMC calculations. Finally, we calculate the densities for the atoms in the LO state and show that the broken translational symmetry shows up.

Now that the lattice geometry has been specified, the dispersion $\varepsilon_{\bf k}$ in Eq.(\ref{hamnd}) and further can also be determined and reads
\begin{align}
\varepsilon_{\bf k}=2t\left[2-\cos(k_xd)-\cos(k_yd)\right],
\label{dispersion}
\end{align}
where $t$ is the hopping parameter and $d$ the lattice spacing. In the following, all energies presented are in units of $t$ and all distances in units of $d$.

In the case we consider in this paper,  where the hopping parameters are the same for the two spin states $t_\uparrow=t_\downarrow\equiv t$, changing the imbalance $h$ to $-h$ only means interchanging the spin labels $\uparrow$ and $\downarrow$. We therefore only consider positive $h>0$, which corresponds to the $\uparrow$ particles being the majority component.

\subsection{Comparing the Fulde-Ferrell and Larkin-Ovchinnikov states}
\begin{figure}
\begin{center}
\includegraphics[scale=.55]{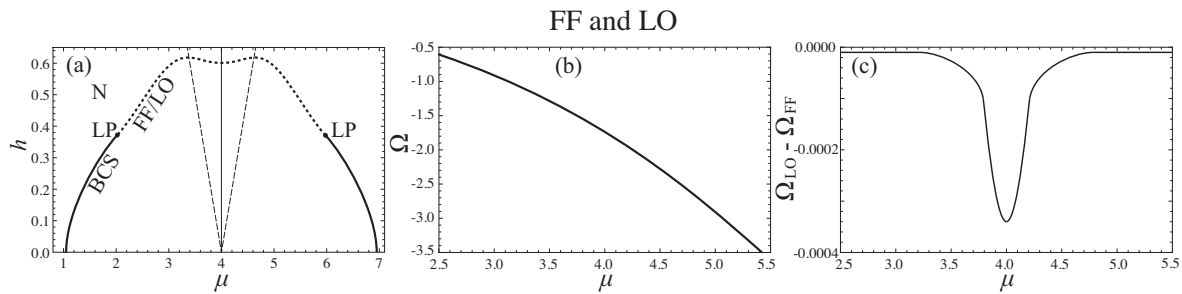}
\caption{(a) The phase diagram for the two-dimensional square lattice at a fixed interaction $U=-3.3$ and temperature $1/k_{\text B}T=5$ as a function of average chemical potential $\mu$ and chemical potential difference $h$. The continuous transition from the normal state (N) to the homogeneous superfluid (BCS) is denoted by a full line and the continuous transition to the FF or LO state by a dashed line, with the Lifshitz point (LP) in between. The dashed lines starting at $\mu=4$ are the Van Hove singularities at zero temperature $\mu_\sigma=4$. (b) The energies $\Omega_\text{FF}$ and $\Omega_\text{LO}$ slightly below the critical imbalance $h'=h_c-0.01$ and (c) the difference in energies $\Omega_\text{LO}-\Omega_\text{FF}$.
}\label{FFandLO}
\end{center}
\end{figure}
We first compare the exotic FF and LO states by calculating for a fixed temperature and interaction strength at different chemical potentials phase transitions from the normal to the superfluid state, the latter being either the BCS state or an FF or LO state. Both the transition lines obtained using $\Omega_\text{FF}$ and $\Omega_{LO}$ are shown in Fig.\ref{FFandLO} as a function of the average chemical potential $\mu=(\mu_\uparrow+\mu_\downarrow)/2$ and the chemical potential difference $h=(\mu_\uparrow-\mu_\downarrow)/2$. 
The transition lines are obtained by minimizing $\Omega_\text{FF}$ and $\Omega_\text{LO}$ with respect to both the Cooper pair amplitude $\Delta_0$ and momentum ${\bf q}$, where we have fixed the direction of the momentum along one of the axis of the lattice, ${\bf q}=q\hat{e}_x$. Namely, when the global minimum of the thermodynamic potential is located at $\Delta_0=0$ the system is in the normal state, whereas a phase transition towards a superfluid phase has occurred when the global minimum is at a nonzero $\Delta_0$. When in the latter case the global minimum of the thermodynamic potential is located at zero center-of-mass momentum, $\Delta_0\neq 0$ and $q=0$, the transition is to a BCS superfluid, whereas it is towards an FF or LO state when the minimum of $\Omega_\text{FF}$ and $\Omega_\text{LO}$ is located at nonzero momentum, $\Delta_0\neq 0$ and $q\neq0$ \cite{PhysRevA.82.013624}. At this point, we only study the instabilities towards superfluid phases, whereas below we also study the superfluid phase itself.

As shown in Fig.\ref{FFandLO}(a), we find for small imbalances $h$ a continuous transition from the normal state to a homogeneous superfluid, where the center-of-mass momentum of the Cooper pairs is zero. For larger imbalances the transition is from a normal state to either an FF or LO superfluid state. The multicritical point where the normal, BCS and FFLO states meet, is called a Lifshitz point and it is marked in the phase diagram (LP).
Above the Lifshitz point, the two different thermodynamic potentials, $\Omega_\text{FF}$ and $\Omega_\text{LO}$, both show a continuous transition towards a superfluid phase at exactly the same chemical potential difference $h_c$. To determine which of the two actually forms the ground state below $h_c$ we have to compare the energies of the two states.

Right at the phase transition, the energy $\Omega_\text{FF}$ is exactly the same as $\Omega_\text{LO}$, since at the critical imbalance $h_c$ they are both equal to the energy of the normal state, $\Omega_\text{N}=\Omega_\text{FF}=\Omega_\text{LO}$. To see which state is more favorable we calculate the energy for imbalances $h'=h_c-0.01$ slightly below the phase transition and the results are shown in Fig.\ref{FFandLO}(b) and (c). In panel (b) the energies $\Omega_\text{FF}$ and $\Omega_\text{LO}$ are both plotted and since they hardly differ it is hard to determine the ground state. However, when the difference between $\Omega_\text{FF}$ and $\Omega_\text{LO}$ is plotted, see Fig.\ref{FFandLO}(c), it is clear that the LO state has a lower energy and thus forms the ground state of the system.   
In conclusion, we find that also in a lattice the LO and not the FF state forms the ground state of the system, which is in agreement with previous work comparing FF and LO in a homogeneous system \cite{2010RPPh...73g6501R,PhysRevB.71.214504,PhysRevA.75.063601,JildouenHenk_PRA2013}.

In the phase diagram Fig.\ref{FFandLO} there are some general features for the two-dimensional square lattice that can also be observed. Firstly, the phase diagram is symmetric on different sides of $\mu=4$, which can be explained by the fact that this value of the average chemical potential corresponds to half filling, see Eq.(\ref{dispersion}). 
Secondly, there are two maximum critical imbalances $h$, one at $\mu$ slightly below and one slightly above half filling. This is due to the nesting of the Fermi surfaces being optimal around the so-called Van Hove singularity, {\it i.e.}, when one of the Fermi surfaces touches the edge of the first Brillouin zone \cite{1367-2630-10-4-045014}. At zero temperature the Van Hove singularities occur for $\mu_\sigma=4$, which are the two dashed lines shown in the phase diagram Fig.\ref{FFandLO}(a).
Because of the symmetry around $\mu=4$, we only consider $\mu<4$ in the rest of the paper.

\subsection{The LO state with different directions}
\begin{figure}
\begin{center}
\includegraphics[scale=.57]{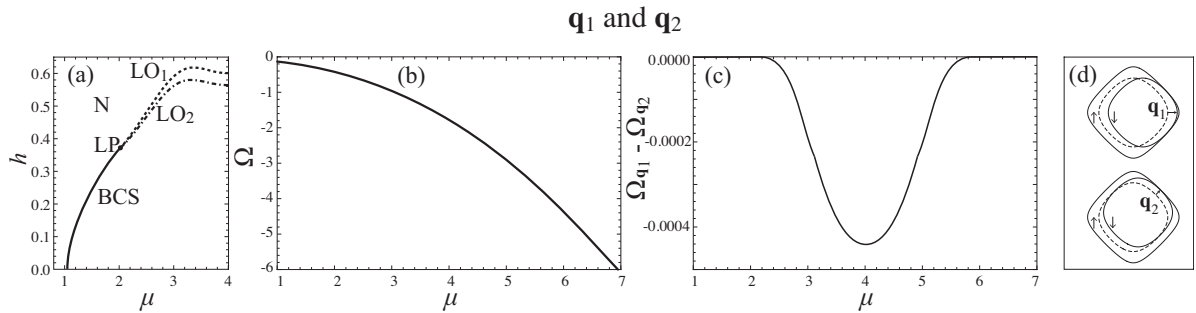}
\caption{(a) The phase diagram at fixed interaction $U=-3.3$ and temperature $1/k_{\text B}T=5$ as a function of $\mu$ and $h$. The transition from the normal state (N) to the BCS state is denoted by a full line and ends in the Lifshitz point (LP). The continuous transition to the LO$_1$ state is denoted by a dashed line and to the LO$_2$ state by a dashed-dotted line. (b) The energies $\Omega_{{\bf q}1}$ and $\Omega_{{\bf q}2}$ at the transition line from normal to LO$_2$ and (c) the difference in energies $\Omega_{{\bf q}1}-\Omega_{{\bf q}2}$. In (d) the Fermi surfaces at zero temperature are shown for the majority ($\uparrow$) and the minority ($\downarrow$) component, where the latter in the upper figure is displaced by ${\bf q}_1$ and in the lower figure by ${\bf q}_2$.}\label{q1q2}
\end{center}
\end{figure}
When comparing the FF and LO superfluid phases we fixed the direction of the center-of-mass momentum of the Cooper pairs for both phases to ${\bf q}=q\hat{e}_x$, along one of the axis of the lattice, and then found that the LO state is energetically more favorable. We now compare LO phases with the Cooper pair momentum in different directions. The two extreme directions in a square lattice for the center-of-mass momentum are along one of the axes of the lattice, here called ${\bf q}_1$, and along the diagonal of the lattice, ${\bf q}_2=q(\hat{e}_x+\hat{e}_y)/\sqrt{2}$, which we therefore consider here.

As before, we calculate the phase diagrams  as functions of average chemical potential $\mu$ and chemical potential difference $h$ for a fixed temperature and interaction strength, see Fig.\ref{q1q2}(a). Below the Lifshitz point the transition is from a normal to the BCS state, the same as in Fig.\ref{FFandLO}(a). Above the Lifshitz point, we find a higher critical imbalance $h_c$ for the momentum ${\bf q}_1$ than for ${\bf q}_2$, the transition lines denoted in Fig.\ref{q1q2}(a) by LO$_1$ and LO$_2$ respectively. This means that for those imbalances it is actually the LO$_1$ state that forms the ground state.

In order to see which inhomogeneous superfluid state occurs below the transition line for LO$_1$, we calculate the energy $\Omega_{{\bf q}1}$ and $\Omega_{{\bf q}2}$ along the transition line LO$_2$ in Fig.\ref{q1q2}(a). The results are shown in Fig.\ref{q1q2}(b) and Fig.\ref{q1q2}(c). Also here it is hard to draw a conclusion from the energy plot. while from the difference in energy it is clear that the LO$_1$ state is energetically more favorable.

This result can be understood from the shape of the Fermi surfaces in a square lattice. In Fig.\ref{q1q2}(d) the zero temperature Fermi surfaces are sketched and it can be seen that when the minority Fermi surface is displaced by ${\bf q_1}$, in the $k_x$-direction, the overlap of the two Fermi surfaces is bigger than in the case of a ${\bf q_2}$ displacement along the diagonal. The overlap is favorable for pairing, which explains why the LO$_2$ state is energetically more favorable.

\subsection{Complete Phase Diagrams}
Now that we have shown that the LO state with Cooper pair momentum ${\bf q}=q\hat{e}_x$ is energetically the most favourable state, we complete the phase diagram by calculating the transition line between the LO and the BCS state. We also calculate the phase diagrams for lower temperatures and for stronger interactions. 

\begin{figure}
\begin{center}
\includegraphics[scale=.82]{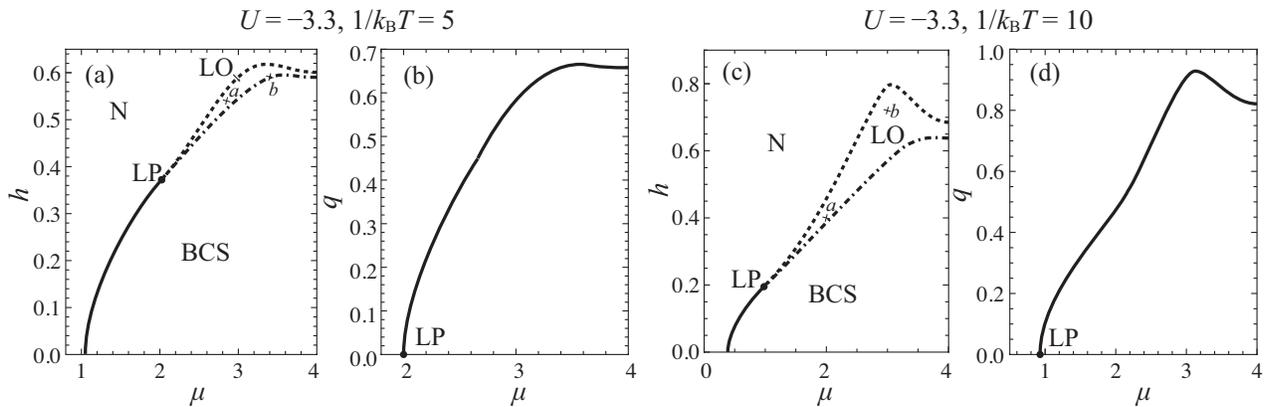}
\caption{Phase diagrams at a fixed interaction $U=-3.3$ as a function of average chemical potential $\mu$ and chemical potential difference $h$ at temperature $1/k_{\text B}T=5$ (a) and $1/k_{\text B}T=10$ (c), containing a normal state (N) and both BCS and LO superfluid phases. These three phases meet at the Lifshitz point (LP). All lines denote continuous transitions. In (b) and (d) the momentum $q$ is plotted for which the transition from normal to LO is found in (a) and (c) respectively. It can be seen that $q$ is more or less proportional to $h$. Particle densities are calculated in Fig.\ref{densbeta5} for the points $a$ and $b$ marked in (a), and in Fig.\ref{densbeta10} for $a$ and $b$ marked in (c).}\label{LOBCSU33}
\end{center}
\end{figure}

The transition line between the LO and the BCS state can be determined from the full thermodynamic potential $\Omega_\text{LO}(\Delta_0,{\bf q}=q\hat{e}_x)$. In the LO state the location of the global minimum of $\Omega_\text{LO}$ is found at both nonzero pairing amplitude $\Delta_0$ and momentum $q$, whereas in the BCS state $\Omega_\text{LO}$ is minimized by a nonzero $\Delta_0$ and zero $q$ \cite{JildouenHenk_PRA2013}. By finding the imbalance where these two minima are the same, we determine the transition line between BCS and LO, see for instance Fig.\ref{LOBCSU33}(a). We find this transition to be a continuous phase transition, in contrast to results obtained with the FF ansatz \cite{PhysRevLett.99.120403,1367-2630-10-4-045014}.

In Fig.\ref{LOBCSU33} the complete phase diagrams are shown for a fixed interaction $U=-3.3$ at temperature $1/k_\text{B}T=5$ in panel (a) and at $1/k_\text{B}T=10$ in panel (c). It can be seen that for lower temperature phase transitions occur at smaller $\mu$ and larger $h$ compared to higher temperature. The Lifshitz point has shifted down and the LO region is considerably larger. 
Also shown in Fig.\ref{LOBCSU33} are the momenta $q$ for which the transitions from the normal state to the LO state are found, in panel (b) for $1/k_\text{B}T=5$ and in panel (d) for $1/k_\text{B}T=10$. By comparing the phase diagrams with these plots, it can be seen that the momentum $q$ is more or less proportional to the chemical potential difference $h$ at the phase transition.

\begin{figure}
\begin{center}
\includegraphics[scale=.82]{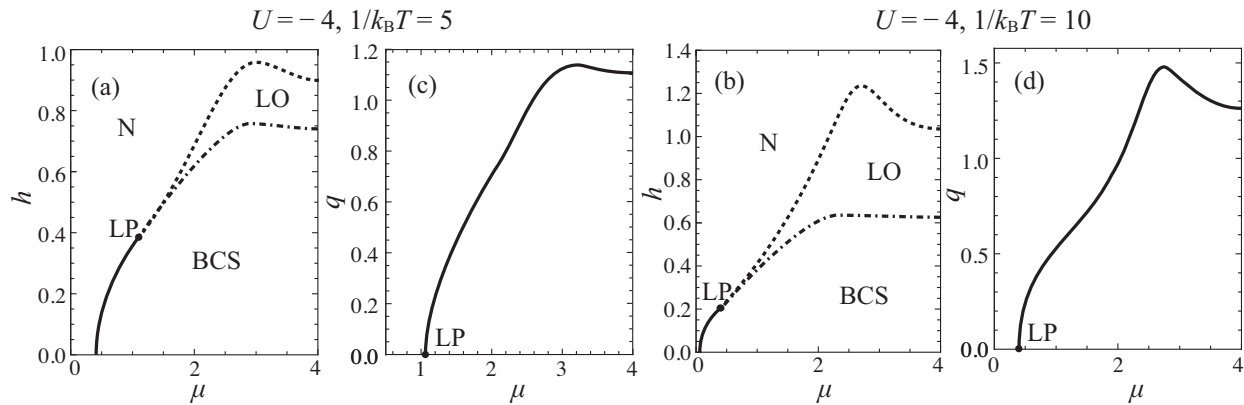}
\caption{Phase diagrams at fixed interaction $U=-4$ as a function of average chemical potential $\mu$ and chemical potential difference $h$ at temperature $1/k_{\text B}T=5$ (a) and $1/k_{\text B}T=10$ (c), containing a normal state (N) and both BCS and LO superfluid phases. These three phases meet at the Lifshitz point (LP). All lines denote continuous transitions. In (b) and (d) the momentum $q$ is plotted for which the transition from normal to LO is found in (a) and (c) respectively. It can be seen that $q$ is more or less proportional to $h$.}\label{LOBCSU4}
\end{center}
\end{figure}

In Fig.\ref{LOBCSU4} the same phase diagrams and momentum plots are shown, but now at a stronger interaction $U=-4$. Again in panel (a) and (b) the temperature is $1/k_\text{B}T=5$ and in panel (c) and (d) $1/k_\text{B}T=10$. Compared to the $U=-3.3$ case all superfluid phase regions are larger. Also for this interaction strength we find all transitions to be continuous.

\subsection{Fixed filling}
Because in experiment the relevant quantities are the particle numbers $n_\sigma$ and not the chemical potentials $\mu_\sigma$, we now calculate the phase diagram at a fixed filling fraction of the lattice as function of temperature and polarization. The particle numbers can be obtained from the thermodynamic potential
\begin{align}
n_\sigma=-\frac{\partial\Omega}{\partial\mu_\sigma}
\end{align}
and in turn the filling fraction $f=(n_\uparrow+n_\downarrow)/2$ and polarization $P=(n_\uparrow-n_\downarrow)/(n_\uparrow+n_\downarrow)$ can be determined.
\begin{figure}
\begin{center}
\includegraphics[scale=.75]{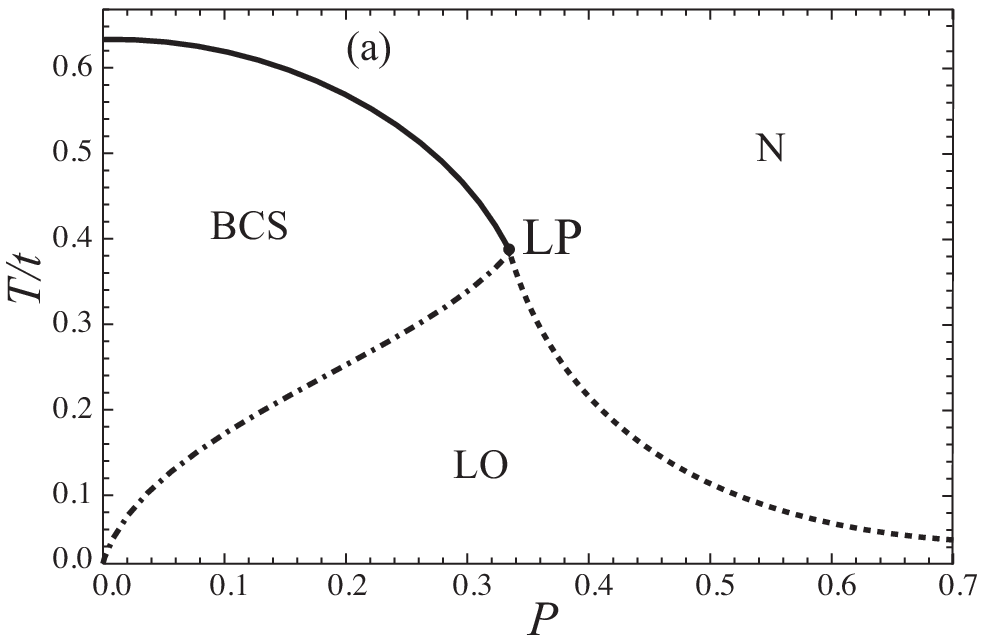}
\includegraphics[scale=.8]{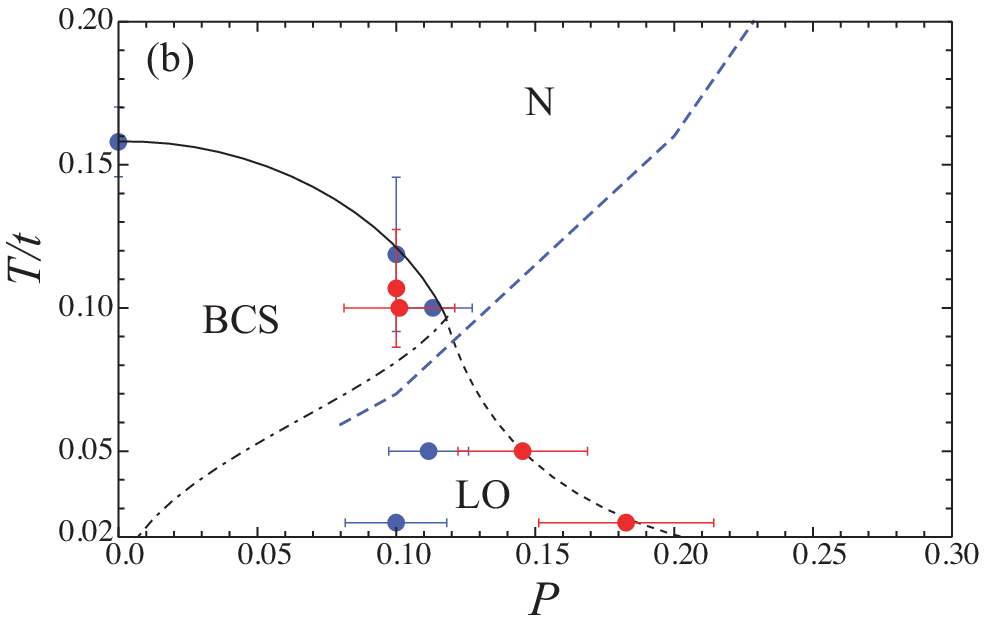}
\caption{(Color online.) (a) Phase diagram at fixed filling $f=0.25$ as function of temperature, scaled by the hopping $T/t$, and polarization $P$ containing a normal phase (N) and both BCS and LO superfluid phases. The three phases meet at the Lifshitz point (LP) and all transition lines correspond to continuous phase transitions. (b) The same phase diagram as a function of temperature $T/t$ and polarization $P$, where our temperature is rescaled by $T'$ and polarization by $P'$ as explained in the main text. Also shown are diagrammatic Monte Carlo results from \cite{DiagMC_Troyer2015}. The blue dots mark the phase boundaries of the BCS superfluid phase and the red dots mark transitions from the normal to an FFLO state. The dashed blue line separates the BCS pairing dominated and the FFLO pairing dominated regions.}\label{fixedf}
\end{center}
\end{figure}

We calculate the phase diagram for a lattice at quarter filling $f=0.25$ as a function of polarization and temperature, see Fig.\ref{fixedf}(a). For the balanced system, $P=0$, the critical temperature is highest and it decreases with increasing polarization. For small polarizations the continuous phase transition is towards a BCS state, while it is to the LO state for larger polarizations and the three phases meet again at the Lifshitz point. Also the transition between the BCS and LO state we find to be continuous. We do not find a region of phase separation, since it would require a first order phase transition. This is in contrast to the fixed filling phase diagram obtained using $\Omega_\text{FF}$, where there is phase separation stemming from the first order phase transition between the FF and BCS states \cite{PhysRevLett.99.120403}.

Fig.\ref{fixedf}(b) shows the same phase diagram, where our temperatures and polarizations are rescaled, together with the diagrammatic Monte Carlo results from \cite{DiagMC_Troyer2015}. We rescaled all the temperatures by $T'=T_0/T_0^*$, with $T_0$ the critical temperature we find at $P=0$, and $T_0^*$ the critical temperature from the Monte Carlo results at $P=0$. All polarizations in our phase diagram are similarly rescaled by $P'=P_0/P_0^*$, with $P_0$ and $P_0^*$ the polarizations at $T=0.02$. Ideally one would use the polarizations at $T=0$ for this rescaling. 
There is an excellent agreement between our rescaled results and the state-of-the-art diagrammatic Monte Carlo results for the transitions from the normal to the BCS or LO state. Remarkably, the Lifshitz point matches very well with the diagMC data using this simple rescaling. This means that qualitatively our mean-field theory seems to capture all relevant physics to map out the phase diagram.

The transition lines separating the LO and the BCS state do not match. Namely, we find this transition to be at smaller polarizations and thus find a bigger LO region than the Monte Carlo results do. In \cite{DiagMC_Troyer2015} also the line that separates the BCS pairing dominated and the FFLO pairing dominated regions is calculated, the dashed line in Fig.\ref{fixedf}(b), and our BCS-LO transition line does follow the trend of that line. 

\subsection{Densities}
As explained before, the LO phase is an inhomogeneous superfluid phase. Namely, translational symmetry is broken in the LO phase in the direction of the Cooper pair momentum ${\bf q}=q\hat{e}_x$, see Eq.(\ref{LOCpair}). This means that the Cooper pair density is position dependent in the $x$-direction, whereas it is homogeneous in the perpendicular direction. In other words, the LO state is a superfluid state, which also has properties of a solid, {\it i.e.}, periodic density modulations. For this reason, the LO phase can be considered to be a special kind of supersolid \cite{PhysRevLett.101.215301}.

Due to the LO symmetry breaking, also the spin densities $n_\sigma$ are position dependent in the $x$-direction, which we calculate here and show that they could serve as a possible experimental signature for the LO phase. To obtain the densities, we need to calculate the coherence factors $u_{{\bf k},n}(x)$ and $v_{{\bf k},n}(x)$ in
\begin{align}
n_{\sigma}(x)=\sum_{{\bf k},n}\left\{u_{{\bf k},n}^2(x)N_\text{F}(\hbar\omega_{{\bf k},\sigma,n})+v_{{\bf k},n}^2(x)\left[1-N_\text{F}(\hbar\omega_{{\bf k},-\sigma,n})\right]\right\},
\end{align}
where ${\bf k}$ lies in the first Brillouin zone and $n$ sums over energy bands  \cite{JildouenHenk_PRA2013}. The Fermi distribution functions are $N_\text{F}(x)=1/[1+\exp(\beta x)]$. The quasiparticle dispersions $\hbar\omega_{{\bf k},n}$ in the above expression are obtained by diagonalizing the $D\times D$ matrix in the LO Hamiltonian ($D$ is varied until convergence). The coherence factors here are similar to the $u_{\bf k,q}$ and $v_{\bf k,q}$ used to diagonalize Eq.(\ref{hambog}) and are obtained by solving the Bogoliubov-de Gennes equation, which we do by expanding
\begin{align}
\left(\begin{array}{c}u_{{\bf k},n}(x)\\ v_{{\bf k},n}(x)\end{array}\right)=\sum_{Q}\left(\begin{array}{c}u_{{\bf k},Q,n}\\ v_{{\bf k},Q,n}\end{array}\right)e^{iQx},
\end{align}
where $Q$ are multiples of the LO wave vector $q$. From this expression it can already be understood that the particle densities have the same wavelength as the LO Cooper pair wave function, namely $\lambda_\text{LO}=2\pi/q$. The factors $u_{{\bf k},Q,n}$ and $v_{{\bf k},Q,n}$ are also obtained from diagonalizing the LO Hamiltonian.

\begin{figure}
\begin{center}
\includegraphics[scale=.85]{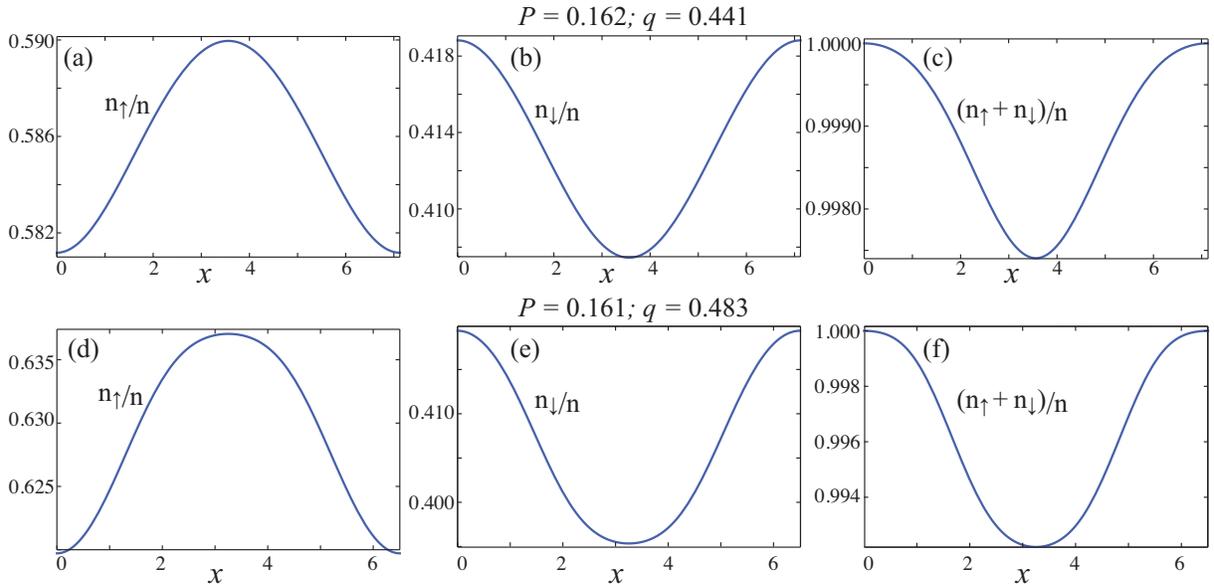}
\caption{The $\uparrow$ particle density, (a) and (d), the $\downarrow$ particle density, (b) and (e), and the total density, (c) and (f), all scaled by $n=n_\uparrow(0)+n_\downarrow(0)$ as a function of position along the $\hat{x}$-direction at interaction $U=-3.3$ and temperature $1/k_{\text B}T=5$. The densities in the top half correspond to point $a$ and in the bottom half to point $b$, both marked in Fig.\ref{LOBCSU33}(a). The polarization $P=0.162$ and LO momentum $q=0.441$ in the top half, $P=0.161$ and $q=0.483$ in the bottom half. The range of the $x$-axis corresponds to the LO wavelength, $\lambda_\text{LO}=2\pi/q$.}\label{densbeta5}
\end{center}
\end{figure}

We calculate the particle densities for several points in the $U=-3.3$ phase diagrams in Fig.\ref{LOBCSU33}. In Fig.\ref{densbeta5}, the densities are shown for the points $a$ and $b$ marked in Fig.\ref{LOBCSU33}(a) at the temperature $1/k_\text{B}T=5$. It can be seen that the density of the minority component $n_\downarrow$ follows the shape of the Cooper pair ansatz, while the majority component $n_\uparrow$ has a maximum where the minority has a minimum. Due to the larger variation in the minority density, the total density $n_\uparrow+n_\downarrow$ has a similar shape to the minority component. Even though the points $a$ and $b$ are not very close to each other in the phase diagram, it turns out that the polarization and Cooper pair momentum are rather similar for the two points,  $P=0.162$ and $q=0.441$ for $a$ and $P=0.161$ and $q=0.483$ for $b$. Yet, the shape of the densities for these two points is rather different, as can be seen by comparing the top and bottom halfs in Fig.\ref{densbeta5}, which is caused by the different Cooper pair amplitudes for the two cases, $\Delta_0=0.217$ in the top and $\Delta_0=0.341$ in the bottom half.

\begin{figure}
\begin{center}
\includegraphics[scale=.85]{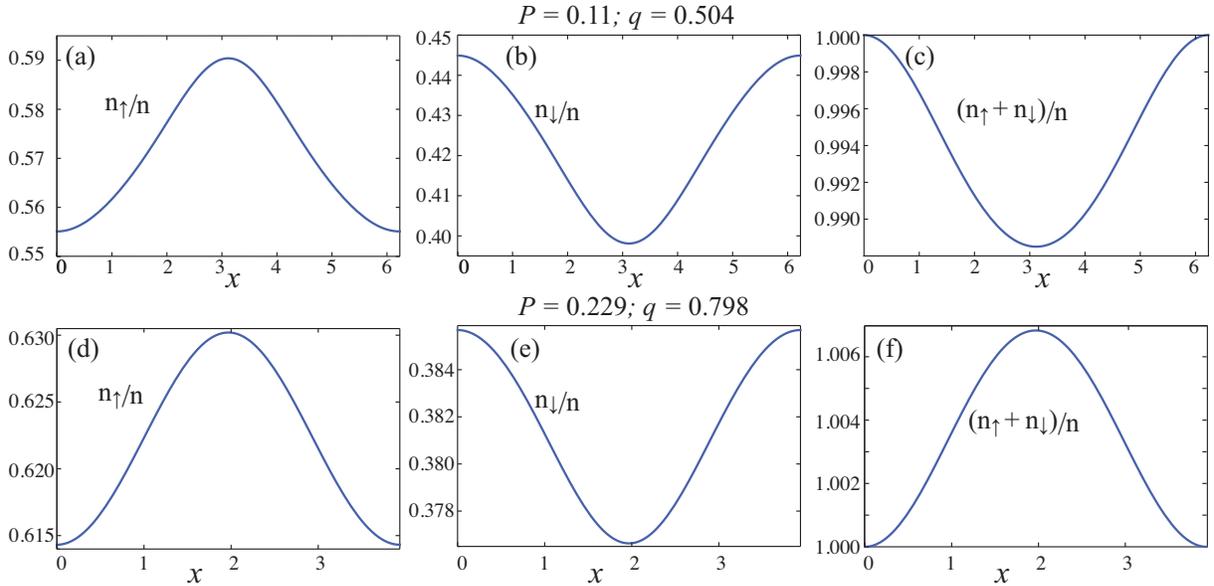}
\caption{The $\uparrow$ particle density, (a) and (d), the $\downarrow$ particle density, (b) and (e), and the total density, (c) and (f), all scaled by $n=n_\uparrow(0)+n_\downarrow(0)$ as a function of position along the $\hat{x}$-direction at interaction $U=-3.3$ and temperature $1/k_{\text B}T=10$. The densities in the top half correspond to point $a$ and in the bottom half to point $b$, both marked in Fig.\ref{LOBCSU33}(c). The polarization $P=0.11$ and LO momentum $q=0.504$ in the top half, $P=0.229$ and $q=0.798$ in the bottom half. The range of the $x$-axis corresponds to the LO wavelength, $\lambda_\text{LO}=2\pi/q$.}\label{densbeta10}
\end{center}
\end{figure}

In Fig.\ref{densbeta10}, the densities are shown for the points $a$ and $b$ marked in Fig.\ref{LOBCSU33}(c) at temperature $1/k_\text{B}T=10$. Comparing with $1/k_\text{B}T=5$, it can be observed that for lower temperature the variations in all the densities are larger. The two points marked in Fig.\ref{LOBCSU33}(c) actually correspond to very different polarizations and Cooper pair momentum, $P=0.11$ and $q=0.504$ for $a$ and $P=0.229$ and $q=0.798$ for $b$. The difference in Cooper pair momentum leads to different wave lengths $\lambda_\text{LO}$, which is clearly visible in the different ranges of the $x$-axis between top and bottom half. The bottom half of Fig.\ref{densbeta10} corresponds to the biggest polarization for which we have calculated densities. The variations in the densities are not the largest for this polarization and, remarkably, the shape of the total density $n_\uparrow+n_\downarrow$ is now the same as the majority component.

From both density figures, Fig.\ref{densbeta10} and Fig.\ref{densbeta5}, it can be seen that the modulations in the atomic densities can be reasonably big, with the largest modulations being around 10$\%$ in Fig.\ref{densbeta10}(a) and (b). It can also be seen that the variations in the total density $n_\uparrow+n_\downarrow$ are  quite small, which is due to the $n_\uparrow$ maximum and $n_\downarrow$ minimum always coinciding. The latter is a general feature of the LO state.

\section{Conclusion and Outlook}
In this paper, we studied Larkin-Ovchinnikov phases in two-component Fermi mixtures in a two-dimensional square lattice. We first showed that in the presence of an imbalance it is actually the LO state, rather than the FF state, that is energetically more favorable. Next, we compared different directions for the center-of-mass momentum and showed that it is directed along one of the lattice axes. Consequently, we calculated phase diagrams at different interactions and temperatures and made a comparison with diagrammatic Monte Carlo results. The good agreement on a qualitative level shows that our mean-field theory captures the relevant physics determining the phase diagram. Lastly, we calculated the particle densities inside the LO state and showed that it has the same dependence on position as the Cooper pairs do.
An interesting extension of our work would be to study the effect of a mass imbalance in the Hubbard model, since it is known to enhance the stability of LO phases in the homogeneous case \cite{PhysRevLett.103.195301,PhysRevA.82.013624}.

\section*{Acknowledgements}
We thank Jan Gukelberger for kindly providing us with their diagrammatic Monte Carlo data \cite{DiagMC_Troyer2015}. This work was supported by the Academy of Finland through its
Centres of Excellence Programme (Projects No. 263347,
No. 251748, and No. 272490) and by the
European Research Council (ERC-2013-AdG-340748-CODE).

\bibliography{FFLO_literature}

\end{document}